\documentstyle[11pt]{article}

\begin{document}
\begin{flushright}
%                     \\
%
December 1996 
\end{flushright}
\vspace{24pt}

\begin{center}
\begin{LARGE}
Degeneration 
of the Elliptic Algebra ${\cal A}_{q,p}\left(\widehat{sl_2}\right)$
and Form Factors in the sine-Gordon Theory
\end{LARGE}

\vspace{30pt}

\begin{large}
Hitoshi Konno\raisebox{2mm}{$\dagger$} 
\end{large}

\vspace{6pt}

\begin{large}
{\it Department of Mathematics,\break
Faculty of Integrated Arts and Sciences,\break 
Hiroshima University,
Higashi-Hiroshima 739, Japan.}

\end{large}

\vspace{35pt}

%\begin{center}
\vspace{35pt}

{ABSTRACT}
\end{center}

\vspace{20pt}
Following the work with Jimbo and Miwa\cite{JKM}, we introduce a certain degeneration of the elliptic algebra ${\cal A}_{q,p}\left(\widehat{sl_2}\right)$ and  its boson realization. We investigate
its rational limit. 
The limit is the central extension of the Yangian double ${\cal D}Y(sl_2)$ 
at level one. We give a new boson realization of it. 
 Based on these algebras, we reformulate the Smirnov's form factor bootstrap 
approach to  the sine-Gordon theory and the $SU(2)$ invariant Thirring model. A conjectural integral formula for form factor in the sine-Gordon theory is derived.

\vspace{22pt}
%\vspace{10pt}
\vfill
\hrule

\vskip 3mm
\begin{small}
\noindent 
Talk given at the Nankai-CRM joint meeting on "Extended and Quantum Algebras and their Applications to Physics", Tianjin, China, August 19-24, 1996. To appear in the CRM series in mathematical physics, Springer Verlag.

\noindent\raisebox{2mm}{$\dagger$} 
 E-mail: konno@mis.hiroshima-u.ac.jp

\end{small}

%--------------------Start of Kyoto definitions --------------------------
%%%%%%%%%%%%%%%%%%%%%%%%%%%%%%%%%%%%%%%%%%%%%%%%%%%%%%%%%%
\renewcommand{\theequation}{\thesection.\arabic{equation}}
%%%%%%%%%%%%%%%%%%%%%%%%%%%%%%%%%%%%%%%%%%%%%%%%%%%%%%%%%%%%%%%%%%%%%%%%%%%

\newcommand{\End}{{\rm End}}
\newcommand{\Res}{{\rm Res}\,}
\newcommand{\id}{{\rm id}}
\newcommand{\sgn}{{\rm sgn}}
\newcommand\sh{{\rm sinh}}
\newcommand{\wt}{{\rm wt}}
\newcommand{\nn}{\nonumber}
\newcommand{\omb}{\underline{\omega}}
\newcommand{\nb}{\underline{n}}
\newcommand{\eqref}[1]{(\ref{#1})}
\newcommand{\refeq}[1]{(\ref{eqn:#1})}
\newcommand{\be}{\begin{equation}}
\newcommand{\en}{\end{equation}}
\newcommand{\bea}{\begin{eqnarray}}
\newcommand{\ena}{\end{eqnarray}}
\newcommand{\bean}{\begin{eqnarray*}}
\newcommand{\enan}{\end{eqnarray*}}
\newcommand{\deq}{\stackrel{\rm def}{=}}
\newcommand{\lb}[1]{\label{eqn:#1}}
\newcommand{\mod}{~\hbox{mod}~}
\newcommand\tr{\hbox{tr}\,}

\newcommand{\Bbb}{\bf}
\newcommand{\Z}{{\Bbb Z}}
\newcommand{\R}{{\Bbb R}}
\newcommand{\C}{{\Bbb C}}
\newcommand{\F}{{\cal F}}
\renewcommand{\O}{{\cal O}}
\newcommand{\ve}{\varepsilon}
\newcommand{\bve}{\bar{\epsilon}}

\def\dlarrow{{\rm h}}
\def\duarrow{{\rm v}}
\newcommand{\lft}{\dlarrow}
\newcommand{\up}{\duarrow}
\newcommand{\br}[1]{{\langle #1 \rangle}}
\newcommand{\ket}[1]{{| #1 \rangle}}
\newcommand{\phit}{\tilde\phi}
\newcommand{\phib}{\bar\phi}

\newcommand{\qed}{\hfill \fbox{}\medskip}
\newcommand{\proof}{\medskip\noindent{\it Proof.}\quad }
\newcommand{\PP}{{\cal P}}
\def\iv#1{{d#1\over2\pi i#1}}
\def\hp{{\hat\pi}}
\def\deg{{\rm deg}}
\def\gl{{\lower.5ex\hbox{$>$}\atop\raise.5ex\hbox{$<$}}}
\def\lg{{\lower.5ex\hbox{$<$}\atop\raise.5ex\hbox{$>$}}}
\def\bX{{\overline X}}
\def\half{{1\over2}}
\def\frac#1#2{{#1\over#2}}
\def\o{\omega}
\def\e{\epsilon}
\def\vep{\varepsilon}
\def\barep{\bar\e}
\def\ket#1{|#1\rangle}
\newtheorem{thm}{Theorem}[section]
\newtheorem{prop}[thm]{Proposition}
\newtheorem{lem}[thm]{Lemma}
\newtheorem{dfn}[thm]{Definition}
\def\cip(#1){(#1;p,q^4)_\infty}
\def\z{\zeta}
\def\qdet{\hbox{$q$-det}}
\def\snh{\hbox{\rm snh}}
\def\Lpm#1{\mathrel{\mathop{\kern0pt L^\pm}\limits^#1}}
\def\L#1{\mathrel{\mathop{\kern0pt L}\limits^#1}}
\def\Lm#1{\mathrel{\mathop{\kern0pt L^-}\limits^#1}}
\def\Lp#1{\mathrel{\mathop{\kern0pt L^+}\limits^#1}}
\def\vep{\varepsilon}
\def\H{{\cal H}}
%%%%%%%%%%%%%%%%%%%%%%%%%%%%%%%%%%%%%%%%%%%%%%%%%%%%%%%%%%
%--------------------End of Kyoto definitions ----------------------------
\setcounter{section}{0}
\setcounter{equation}{0}

\section{Introduction}\label{sec:1}

Recently,  a representation theory of the 
quantum affine algebra $U_q(\widehat{sl_2})$  has been studied 
extensively and applied successfully to the $XXZ$ model in the anti-ferromagnetic regime\cite{JM}. 
There the $R-$matrix $\left(R_{\vep_1\vep_2}^{\vep_1'\vep_2'}(\z)\right),\ \vep_j,\vep_j'=\pm\ j=1,2$ associated with the $XXZ$ model 
is identified with the  intertwiner of the tensor product space of the two-dimensional evaluation modules $V_{\z}$, and the  two level one infinite-dimensional highest weight modules 
${\cal H}^{(i)}, \ i=0,1$ yield the exact construction of the doubly degenerated physical space of states in the thermodynamic limit. Moreover, there exist two types of  intertwining operators, called type I
and type II vertex operator,
 of the form
\begin{eqnarray*}
{\rm Type\  I}\quad &&\Phi^{(1-i,i)}(\z):\H^{(i)}\longrightarrow \H^{(1-i)}\otimes V_\z,
\\
&&\Phi^{(1-i,i)}(\z)=\sum \Phi^{(1-i,i)}_\vep(\z)\otimes v_\vep,
\\
{\rm Type\  II}\quad&& \Psi^{*(1-i,i)}(\z):V_\z\otimes \H^{(i)}\longrightarrow \H^{(1-i)},
\\
&&\Psi^{*(1-i,i)}_\vep(\z)=\Psi^{*(1-i,i)}(\z)\left(v_\vep\otimes\cdot\right).
\end{eqnarray*}
Each type of vertex operator plays very different role.
The type I vertex operators in their certain combination give the embedding of  spin operators sitting on the lattice into the physical space of states, whereas the type II vertex operator  creates one  physical excited particle. Hence the use of the two types of vertex operators enables us to calculate the spin-spin correlation functions as well as the form factors of the spin operators.
Remarkably, the whole properties of these vertices are summarized
in the following simple relations. 
%\begin{enumerate}
%\item 

\noindent 
{ \it 1. Commutation relations}
\begin{eqnarray}
\Phi_{\vep_2}(\z_2)\Phi_{\vep_1}(\z_1)
&=&
\sum_{\vep_1',\vep_2'=\pm} R_{\vep_1\vep_2}^{\vep_1'\vep_2'}(\z_1/\z_2)
\Phi_{\vep_1'}(\z_1)\Phi_{\vep_2'}(\z_2),
\label{eqn:PhiPhi}
\\
\Phi_{\vep_1}(\z_1)\Psi^*_{\vep_2}(\z_2)
&=&
{\cal T}(\z_1/\z_2)\Psi^*_{\vep_2}(\z_2)\Phi_{\vep_1}(\z_1),
\label{eqn:PhiPsi}
\\
\Psi^*_{\vep_1}(\z_1)\Psi^*_{\vep_2}(\z_2)
&=&
\sum_{\vep_1',\vep_2'=\pm} S^{\vep_1'\vep_2'}_{\vep_1\vep_2}(\z_1/\z_2)
\Psi^*_{\vep_2'}(\z_2)\Psi^*_{\vep_1'}(\z_1),
\label{eqn:PsiPsi}
\end{eqnarray}
where $\left( S^{\vep_1'\vep_2'}_{\vep_1\vep_2}(\z)\right)$ is the two-body $S-$matrix of the physical excited particles and ${\cal T}(\z)$ is a certain scalar function.

\noindent
{\it  2. Normalization conditions}
\begin{eqnarray}
&&\Phi^{(i,1-i)}_{\vep_1}(\z_1)\Phi^{(1-i,i)}_{\vep_2}(\z_2)
=(-1)^{1-i}\vep_2g^{-1}\delta_{\vep_1+\vep_2,0}+O(\zeta_1-q\zeta_2)
\quad (\zeta_1\rightarrow q\zeta_2),
\nonumber\\
&&\qquad \quad\label{eqn:nor1}\\
&&\Psi^{*(i,1-i)}_{\vep_1}(\z_1)\Psi^{*(1-i,i)}_{\vep_2}(\z_2)
=\frac{(-1)^{1-i}\vep_1}{1-q^{-2}\zeta_2^2/\zeta_1^2}
\left(\frac{\zeta_2}{q\zeta_1}\right)^{i+(1+\vep_1)/2}g\delta_{\vep_1+\vep_2,0}
+O(1)
\nonumber\\
&&\qquad\quad (\zeta_1\rightarrow q^{-1}\zeta_2),
\label{eqn:nor2}
\end{eqnarray}
with some constant $g$.
%\end{enumerate}

%Using these properties, one can easily derive a difference equation, %called $q-$deformation of the Knizhnik-Zamolodchikov equation ($q-%$KZE), 
%as the equation satisfied by the correlation functions and form factors %of local spin-operators\cite{JM}.

On the contrary to 
the solvable lattice models,  algebraic study of the 
massive integrable theories has not yet been so much developed. For the on-shell $S-$matrix, we know that the Zamolodchikov's bootstrap approach\cite{Zam}, a scheme of calculation of the $S-$matrix, was reformulated as an algebraic problem in the representation theory  of  quantum groups\cite{Jimbo,Bernard}.  This was  done in the analogous way  to the $R-$matrix in the lattice models. On the other hand, 
for the off-shell quantities such as form factors of some local operators, we know Smirnov's bootstrap approach as a well-founded scheme of calculation\cite{Smi92}.
Let $f(\beta_1,..,\beta_n)_{\epsilon_1,..\epsilon_n}$ be a form factor 
of some local operator with rapidities $\beta_j$ and spins $\vep_j, j=1,2,..,n$. 
His approach is based on the following three axioms.

%\begin{description}
\noindent
{\it Axiom 1}\quad  The S-matrix symmetry
\begin{eqnarray} 
&&f(\beta_1,..,\beta_{i}\beta_{i+1},.
.,\beta_n)_{\epsilon_1,..,\epsilon_{i}\epsilon_{i+1},..,\epsilon_n}
S^{\epsilon'_i\epsilon'_{i+1}}_{\epsilon_i\epsilon_{i+1}}(\beta_i-\beta_{i+1}) 
 \nonumber \\
&& =f(\beta_1,..,\beta_{i+1}\beta_{i},.
.,\beta_n)_{\epsilon_1,..,\epsilon'_{i+1}\epsilon'_{i},..,\epsilon_n}.
\label{axiom1}
\end {eqnarray}

\noindent
{\it Axiom 2}\quad 
\begin{equation} 
f(\beta_1,..,\beta_n+2\pi i)_{\epsilon_1,.,\epsilon_n}
=f(\beta_n,\beta_1,..,\beta_{n-1})_{\epsilon_n,\epsilon_{1},..,\epsilon_{n-1}}.
\label{axiom2}
\end{equation}

\noindent
{\it Axiom 3}\quad As a function of $\beta_n$, form factor 
$f(\beta_1,..,\beta_n)_{\epsilon_1,..\epsilon_n}$ is an analytic function in the strip $0\leq {\rm Im}\beta_n\leq 2\pi$ and has  only simple poles at 
$\beta_n=\beta_j+\pi i,\quad n>j$. The  corresponding 
residues are given by
\begin{eqnarray} 
&&2\pi i\ {{\rm res}}\
f(\beta_1,..,\beta_n)_{\epsilon_1,..,\epsilon_n}
\nonumber \\
&&=f(\beta_1,..,\hat{\beta_j},..,\beta_{n-1})_{\epsilon'_1,..,
\hat{\epsilon}'_j,..,\epsilon'_{n-1}}
{\cal C}_{\epsilon_n,\epsilon'_{j}}
 \nonumber \\
&& \times \Bigl(
\delta^{\epsilon'_1}_{\epsilon_1}\cdots
\delta^{\epsilon_{j-1}'}_{\epsilon_{j-1}}
S^{\epsilon'_{n-1}\epsilon'_{j}}_{\epsilon_{n-1}\tau_1}(\beta_{n-1}-\beta_{j}) 
S^{\epsilon'_{n-2}\tau_{1}}_{\epsilon_{n-2}\tau_2}(\beta_{n-2}-\beta_{j}) 
\cdots S^{\epsilon'_{j+1}\tau_{n-j-2}}_{\epsilon_{j+1}\epsilon_{j}}
(\beta_{j+1}-\beta_{j})\nonumber\\
&&\qquad-S^{\epsilon'_{j}\epsilon'_{1}}_{\tau_1\epsilon_{1}}(\beta_{j}-\beta_{1}) 
\cdots S^{\tau_{j-3}\epsilon'_{j-2}}_{\tau_{j-2}\epsilon_{j-2}}
(\beta_{j}-\beta_{j-2}) S^{\tau_{j-2}\epsilon'_{j-1}}_{\epsilon_{j}\epsilon_{j-1}}
(\beta_{j}-\beta_{j-1})\delta^{\epsilon'_{j+1}}_{\epsilon_{j+1}}\cdots
\delta^{\epsilon_{n-1}'}_{\epsilon_{n-1}}
\Bigr),\nonumber\\
\label{axiom3}
\end{eqnarray}
where ${\cal C}$ is the charge conjugation matrix.
%\end{description}

 These axioms define a matrix Riemann-Hilbert problem
for the functions $f(\beta_1,..,\beta_n)_{\epsilon_1,..\epsilon_n}$.
Remarkably, quite similar properties to (\ref{axiom1})-(\ref{axiom3}) are satisfied by the correlation functions and form factors in the lattice models (See for example \cite{JM96}). Hence one may expect that {\it there must be some
 algebraic formulation of the Smirnov's form factor bootstrap approach}.  

In fact,  in \cite{Smir}, Smirnov considered the $SU(2)$ invariant Thirring model, which is a certain limit of the sine-Gordon theory, and proposed that the Yangian double ${\cal D}Y(sl_2)$ is a relevant algebra. He conjectured also that the level-0 representation is a relevant representation\cite{Smir}.  However due to  lack of infinite dimensional representation, he failed to reproduce his whole axioms.  

The second  progress was brought by Lukyanov\cite{Luk95}.
He considered  the sine-Gordon theory and investigated
an algebra of vertex operators (\ref{eqn:ZZ1})-(\ref{eqn:ZZ3})
which are very similar to (\ref{eqn:PhiPhi})-(\ref{eqn:nor2}). 
 He showed that such algebra formally reproduces Smirnov's whole axioms.  
%Therefore a proper representation of the algebra allows  
%a calculation of form factor of the theory. 
%He  proposed a boson representation of the algebra.  
%However his representation  depends on the cut-off parameter and only %after the limit cut-off going to zero, one obtains correct relations for %the vertex operators. Hence an identification of his result with any %quantum groups had not  been done untill Ref.\cite{JKM}.  
However, his argument is rather a phenomenological one. No one had 
succeeded in giving  it any representation theoretical foundation
untill Ref.\cite{JKM}.

In this paper, according to the prior work \cite{JKM}, 
we clarify an underlying quantum group structure of 
Smirnov's form factor bootstrap and Lukyanov's algebra.
We introduce a
certain degeneration of the elliptic algebra ${\cal A}_{q,p}(\widehat{sl_2})$\cite{FIJKMY} and present a boson representation of it. 
We investigate also its rational limit and identify it with 
the central extension of the Yangian double ${\cal D}Y(sl_2)$ at level one. 
We show that in the both cases certain gauge transformations of 
the type I and the type II vertex operators satisfy the Lukyanov's algebra.
As a result, we give formulae for form factors, 
in the case of the sine-Gordon theory and the $SU(2)$ invariant Thirring model, which satisfy  the whole Smirnov's axioms, based on 
the representations of our algebras. 
This is natural because the sine-Gordon theory is known as an continume
limit of the $XYZ$ model\cite{Luth}, and the elliptic algebra
${\cal A}_{q,p}(\widehat{sl_2})$ is the algebra which is conjectured to 
give an algebraic foundation of the $XYZ$ model ( See for example \cite{JKKMW}).

We hence obtain a unified way, whose key formulae are given by
(\ref{eqn:PhiPhi})-(\ref{eqn:nor2}),
 to formulate correlation functions and form factors both in the exactly solvable lattice models and the massive
integrable quantum field theories.
% conclude that an relevant algebra for the sine-Gordon theory is 
%the degeneration of the elliptic algebra ${\cal A}_{q,p}(\widehat{sl_2})$ 
%whereas the one relevant to  
%the $SU(2)$ invariant Thirring model is ${\cal D}Y(sl_2)_1$. 
%The latter result is consistent to the work in Ref.\cite{KLP}.
We finally give a conjectural integral formula for form factor in the sine-Gordon theory.

\setcounter{section}{1}
\setcounter{equation}{0}

\section{Elliptic algebra ${\cal A}_{q,p}\left(\widehat{sl_2}\right)$ } 
Let us begin with 
the Baxter's elliptic $R$ matrix\cite{Bax82}. 
\begin{equation}
R(\z)=R(\z;p^{1/2},q^{1/2})=
\frac{1}{\mu(\z)}
\pmatrix{a(u)&&&d(u)\cr
&b(u)&c(u)&\cr
&c(u)&b(u)&\cr
d(u)&&&a(u)\cr},\lb{xyzR}
\end{equation}
\begin{equation}
a(u)={\snh(\lambda-u)\over\snh(\lambda)},\quad
b(u)={\snh(u)\over\snh(\lambda)},\quad
c(u)=1,\quad
d(u)=k\,\snh(\lambda-u)\snh(u),
\label{eqn:Boltz}
\end{equation}
where $\snh(u)=-i {\rm sn}(iu)$, and 
${\rm sn}(u)$ is Jacobi's elliptic function with modulus $k$. 
Let $K,K'$ be the corresponding complete elliptic integrals. 
We  use also the variables 
\be
p=e^{-{\pi K'\over K}},
\qquad q=-e^{-{\pi\lambda\over2K}}, 
\qquad\z=e^{{\pi u\over2K}},
\label{addp}
\en
and regard \eqref{eqn:Boltz} as functions of $\z,p,q$. 
We choose the overall scalar factor $\mu(\z)$ as follows. 
\begin{eqnarray}
&&{1\over\mu(\z)}=\frac{1}{\overline{\kappa}(\z^2)}
{(p^2;p^2)_\infty\over(p;p)^2_\infty}
{\Theta_{p^2}(q^2)\Theta_{p^2}(p\z^2)\over\Theta_{p^2}(q^2\z^2)},
\\
&&\frac{1}{\overline{\kappa}(z)}
={\cip(q^4z^{-1})\cip(q^2z)\cip(pz^{-1})\cip(pq^2z)
\over\cip(q^4z)\cip(q^2z^{-1})\cip(pz)\cip(pq^2z^{-1})},
\nonumber
\end{eqnarray}
where 
\begin{eqnarray*}
(z;p_1,\cdots,p_m)_\infty
&=&\prod_{n_1,\cdots,n_m\ge 0}(1-zp_1^{n_1}\cdots p_m^{n_m}),
\\
\Theta_q(z)&=&(z;q)_\infty(qz^{-1};q)_\infty(q;q)_\infty. 
\end{eqnarray*}

%\subsection{Definition of the elliptic algebra  ${\cal A}_{q,p}\left%(\widehat{sl}_2\right)$}
 Let us consider the formal generating series 
\begin{eqnarray}
&&L^\pm(\z)=\sum_{n=-\infty}^\infty L^\pm_n \,\z^{-n},
\qquad
L^\pm_n=\left(L^\pm_{\vep\vep',n}\right)_{\vep,\vep'=\pm}, 
\label{eqn:Lpm1}\\
&&L^\pm_{\vep\vep',n}=0 \quad \hbox{ if  } \vep\vep'\neq (-1)^n.
\nonumber
\end{eqnarray}

\vspace{2mm}
\newpage

\noindent
{\bf Definition :}\quad {\bf the elliptic algebra  ${\cal A}_{q,p}\left(\widehat{sl_2}\right)$}\cite{FIJKMY,hwm}

{\it
The elliptic algebra ${\cal A}_{q,p}(\widehat{sl_2})$ is 
the algebra generated by the symbols 
$L^\pm_{\vep\vep',n}$ $(n\in\Z, \vep,\vep'=\pm, \vep\vep'=(-1)^n)$ 
and a central element $c$, 
through the following relations. 
\begin{eqnarray}
&&R^\pm_{12}(\zeta_1/\zeta_2)\Lpm{1}(\zeta_1)\Lpm{2}(\zeta_2)
=\Lpm{2}(\zeta_2)\Lpm{1}(\zeta_1)R^{*\pm}_{12}(\zeta_1/\zeta_2),
\label{eqn:Rpm1}\\
&&R^+_{12}(q^{c/2}\zeta_1/\zeta_2)\Lp{1}(\zeta_1)\Lm{2}(\zeta_2)
=\Lm{2}(\zeta_2)\Lp{1}(\zeta_1)R^{*+}_{12}(q^{-c/2}\zeta_1/\zeta_2), 
\nonumber\\
&&\label{eqn:Rpm2}\\
&&\qdet L^+(\z)\equiv
L^+_{++}(q^{-1}\z)L^+_{--}(\z)-L^+_{-+}(q^{-1}\z)L^+_{+-}(\z)=q^{c/2},
\label{eqn:qdet}\\
&&L^-_{\vep\vep'}(\z)=\vep\vep' L^+_{-\vep,-\vep'}(p^{1/2}q^{-c/2}\z),
\label{eqn:Lsym}
\end{eqnarray}
where 
\[
R^+(\z)=\tau(q^{1/2}\z^{-1})R(\z),
\qquad
R^-(\z)=\tau(q^{1/2}\z)^{-1}R(\z) 
\]
with
\begin{equation}
\tau(\z)=\z^{-1}
\frac{(q\z^2;q^4)_\infty(q^3\z^{-2};q^4)_\infty}
{(q^3\z^2;q^4)_\infty(q\z^{-2};q^4)_\infty}
\label{eqn:tau}
\end{equation}
and}
\bea
R^{*\pm}(\z)=R^{\pm}(\z;p^{*1/2},q^{1/2}),
\qquad p^*=pq^{-2c}.
\label{eqn:Rstar}
\ena

In \cite{FIJKMY},
it was conjectured  that  the elliptic algebra 
${\cal A}_{q,p}(\widehat{sl_2})$
has  natural analogs of the level one modules $\H^{(i)}$ i=0,1\footnote{
We say that a representation of 
${\cal A}_{q,p}(\widehat{sl}_2)$ 
has level $k$ 
if the central element $c$ acts as $k$ times the identity.
} 
and vertex operators $\Phi^{(1-i,i)}_{\varepsilon}(\zeta)$ and 
$\Psi^{*(1-i,i)}_{\varepsilon}(\zeta)$.
%\footnote{We shall often omit writing the upper indices $i$ when 
%there is no fear of confusion. 
%}. 
It was also  conjectured that these vertex operators satisfy the commutation relation (\ref{eqn:PhiPhi})-(\ref{eqn:nor2}) with the elliptic $R-$matrix (\ref{eqn:xyzR}), 
$S_{\varepsilon_1,\varepsilon_2}^{\varepsilon'_1,\varepsilon'_2}(\zeta)=
-R_{\varepsilon_1,\varepsilon_2}^{*\varepsilon'_1,\varepsilon'_2}(\zeta)$, ${\cal T}(\zeta)=\tau(\zeta)$ and a different constant $g$. 

In terms of the vertex operators, 
the $L^\pm$ operators acting on $\H^{(i)}$ can be expressed as 
follows.
\begin{eqnarray}
L^+_{\vep\vep'}(\z)&=&\kappa \Psi^*_{\vep'}(\z)\Phi_\vep(q^{1/2}\z),
\label{eqn:Miki1}
\\
L^-_{\vep\vep'}(\z)&=&\kappa \Phi_{\vep}(\z)\Psi^*_{\vep'}(q^{1/2}\z).
\label{eqn:Miki2}
\end{eqnarray}
Here $\kappa$ is a normalization constant. 
Then the  defining relations \eqref{eqn:Rpm1},\eqref{eqn:Rpm2} 
are immediate consequences of the elliptic analogue of the commutation relations \eqref{eqn:PhiPhi}
-\eqref{eqn:PsiPsi}. 
The condition \eqref{eqn:qdet}
for the quantum determinant also follows from
\eqref{eqn:nor1},\eqref{eqn:nor2} with an appropriate choice of $\kappa$. 
The symmetry \eqref{eqn:Lsym} of the $L$ operators entails the 
following relation for the vertex operators.
\begin{equation}
\Phi_{\vep}(\z)\Psi^*_{\vep'}(q^{1/2}\z)
=
\vep\vep' \Psi^*_{-\vep'}(p^{1/2}q^{-1/2}\z)\Phi_{-\vep}(p^{1/2}\z).
\label{eqn:Lsym2}
\end{equation}

\setcounter{section}{2}
\setcounter{equation}{0}

\section{Degeneration of ${\cal A}_{q,p}(\widehat{sl_2})$}

\subsection{Trigonometric limit}

There are two interesting degeneration limit. 

\vspace{2mm}
\noindent
1) $K\to \frac{\pi}{2}, K'\to \infty$ i.e. $p\rightarrow 0,\ q\to -e^{-\lambda}$.

Let us  set 
\[
L^+_{\vep\vep',n}=\left(-p^{1/2}\right)^{\max(n,0)}
\overline{L}^+_{\vep\vep',n}
\]
and let formally $p\rightarrow 0$, then 
$L^+(\z)$ and $L^-(\z)$ become power series in $\z$ and $\z^{-1}$ 
respectively. 
In this limit, the relations \eqref{eqn:Rpm1}-\eqref{eqn:Lsym} 
reduce to the defining relations of the 
quantum affine algebra $U_q(\widehat{sl}_2)$ due to 
Reshetikhin and Semenov-Tian-Shanskii \cite{RS}.

\vspace{2mm}
\noindent
2) $K\to \infty, K'\to \frac{\pi}{2}$. 
More precisely, we let\cite{JKM}
\begin{equation}
p=q^{2(\xi+1)},\quad \z=q^{i\beta/\pi},
\qquad q\rightarrow 1
\label{eqn:lim}
\end{equation}
with $\xi$ and $\beta$ being kept fixed. 

In this limit, the elliptic 
$R$ matrices \eqref{eqn:xyzR} and $R^*(\zeta)=R(\zeta;p^{*1/2};q^{1/2})$
 degenerate to  trigonometric ones. 
\bea
\tilde{R}^*(\beta)&=&\lim R(\zeta;p^{*1/2};q^{1/2})\nonumber \\
&=&-S_0(\beta)\pmatrix{
\frac{\cosh\frac{i\pi}{2\xi}\cosh\frac{\beta}{2\xi}}
{\cosh \frac{i\pi-\beta}{2\xi}}&&&
-\frac{\sinh\frac{i\pi}{2\xi}\sinh\frac{\beta}{2\xi}}
{\cosh \frac{i\pi-\beta}{2\xi}}\cr
&-\frac{\cosh\frac{i\pi}{2\xi}\sinh\frac{\beta}{2\xi}}
{\sinh \frac{i\pi-\beta}{2\xi}}&\frac{\sinh\frac{i\pi}{2\xi}\cosh\frac{\beta}{2\xi}}{\sinh \frac{i\pi-\beta}{2\xi}}&\cr
&\frac{\sinh\frac{i\pi}{2\xi}\cosh\frac{\beta}{2\xi}}{\sinh \frac{i\pi-\beta}{2\xi}}&-\frac{\cosh\frac{i\pi}{2\xi}\sinh\frac{\beta}{2\xi}}{\sinh \frac{i\pi-\beta}{2\xi}}&\cr
-\frac{\sinh\frac{i\pi}{2\xi}\sinh\frac{\beta}{2\xi}}{\cosh \frac{i\pi-\beta}{2\xi}}&&&\frac{\cosh\frac{i\pi}{2\xi}\cosh\frac{\beta}{2\xi}}{\cosh \frac{i\pi-\beta}{2\xi}}
\cr}
\nonumber \\
\lb{tildeRstar}\\
\tilde{R}(\beta)&=&\lim R(\zeta;p^{1/2};q^{1/2})\nonumber \\
&=&-\tilde{R}^*(-\beta)|_{\xi\to \xi+1}.
\ena
Here $S_0(\beta)$ is given by
\bea
&&S_0(\beta)=\frac{S_2(-i\beta)S_2(\pi+i\beta)}
{S_2(i\beta)S_2(\pi-i\beta)}
\ena
with $S_2(x)$ being Barnes' double sine function with periods
$2\pi$ and $\pi\xi$\cite{JM96}.
These are not the standard trigonometric 
$R$ matrix coming from the universal $R$ matrix of 
$U_q(\widehat{sl}_2)$. 
In order to bring them to the usual form, 
we need to introduce a `gauge' transformation\cite{JKKMW}.  
Define 
\[
U=\pmatrix{1&-i\cr 1& i\cr},
\qquad 
U_0= U \,\sigma^z,\quad 
U_1=\sigma^z\, U \,\sigma^z. 
\]
Then we find 
\begin{eqnarray}
&&\left(U_1\otimes U_0\right)
\tilde{ R}(\beta)
\left(U_0\otimes U_1\right)^{-1}
=R(-\beta), \label{eqn:gauge1}\\
&&\left(U_1\otimes U_0\right)
\left(-\tilde{R}^*(\beta)\right)
\left(U_0\otimes U_1\right)^{-1}
=S(\beta) 
\label{eqn:gauge2}
\ena
with
\begin{eqnarray}
&&S(\beta)=\frac{S_0(\beta)}{\sinh \frac{i\pi-\beta}{\xi}}
\pmatrix{\sinh \frac{i\pi-\beta}{\xi}&&&\cr
&\sinh\frac{\beta}{\xi}&\sinh\frac{i\pi}{\xi}&\cr
&\sinh\frac{i\pi}{\xi}&\sinh\frac{\beta}{\xi}&\cr
&&&\sinh \frac{i\pi-\beta}{\xi}\cr},
\\
&&R(\beta)=-S(-\beta)\bigl|_{\xi\rightarrow \xi+1}.
\end{eqnarray}
We  have also 
\bea
&&\left(U_0\otimes U_1\right)
\tilde{R}(\beta)
\left(U_1\otimes U_0\right)^{-1}
=R(-\beta), \label{eqn:gauge1*}\\
&& \left(U_0\otimes U_1\right)
\left(-\tilde{R}^{*}(\beta)\right)
\left(U_1\otimes U_0\right)^{-1}
=S(\beta).\label{eqn:gauge2*}
\end{eqnarray}
The matrices $\left(S^{cd}_{ab}(\beta)\right)$, 
$\left(R^{cd}_{ab}(\beta)\right)$ coincide
with the two-body  $S-$matrix of the sine-Gordon theory\cite{Zam} and the $R-$matrix
of the $XXZ$ model in the gapless regime\cite{JM96}.

Now we are interested in the degeneration limit of 
the elliptic analogue of the relations 
\eqref{eqn:PhiPhi}-\eqref{eqn:PsiPsi} .
Write $\Phi_a(\beta),\Psi^*_a(\beta)$ for the limit of
$\Phi_a(\z),\Psi^*_a(\z)$.
We set 
\begin{eqnarray}
&&Z^{(1,0)}_a(\beta)=2\sum_{b} (U_0^{-1})_{ba}\Psi^{*(1,0)}_b(\beta)
=\Psi^{*(1,0)}_+(\beta)-ia\Psi^{*(1,0)}_-(\beta),
\\
&&Z^{(0,1)}_a(\beta)=2\sum_{b} (U_1^{-1})_{ba}\Psi^{*(0,1)}_b(\beta)
=a\Psi^{*(0,1)}_+(\beta)-i\Psi^{*(0,1)}_-(\beta),
\\
&&Z^{'(1,0)}_a(\beta)=\sum_{b} (U_0)_{ab}\Phi^{(1,0)}_b(\beta)
=\Phi^{(1,0)}_+(\beta)+ia\Phi^{(1,0)}_-(\beta),
\\
&&Z^{'(0,1)}_a(\beta)=\sum_{b} (U_1)_{ab}\Phi^{(0,1)}_b(\beta)
=a\Phi^{(0,1)}_+(\beta)+i\Phi^{(0,1)}_-(\beta).
\end{eqnarray}
Their commutation relations can be determined 
using \eqref{eqn:gauge1} and \eqref{eqn:gauge2}.
Dropping the upper indices, we find 
\begin{eqnarray}
Z_a(\beta_1)Z_b(\beta_2)&=&
\sum_{c,d}S^{cd}_{ab}(\beta_1-\beta_2) Z_d(\beta_2)Z_c(\beta_1),
\label{eqn:ZZ1}\\
Z'_a(\beta_1)Z'_b(\beta_2)&=&
\sum_{c,d}R^{cd}_{ab}(\beta_1-\beta_2) Z'_d(\beta_2)Z'_c(\beta_1),
\label{eqn:ZZ2}\\
Z_a(\beta_1)Z'_b(\beta_2)&=&
ab\tan\left(\frac{\pi}{4}+i\frac{\beta_1-\beta_2}{2}\right)
Z'_b(\beta_2)Z_a(\beta_1).
\label{eqn:ZZ3}
\end{eqnarray}
Here we have used 
\begin{equation}
\lim \tau(q^{i\beta/\pi})= 
\tan\left(\frac{\pi}{4}+\frac{i\beta}{2}\right).
\label{eqn:taul}
\end{equation}
The conditions corresponding to \eqref{eqn:nor1} and 
\eqref{eqn:nor2} become \begin{eqnarray}
Z_a(\beta_1)Z_b(\beta_2)&=&
\frac{C}{\beta_1-\beta_2-\pi i}\delta_{a+b,0}+O(1)
\qquad (\beta_1\rightarrow \beta_2+\pi i),
\nonumber\\
&&\label{eqn:no1}\\
Z'_a(\beta)Z'_b(\beta+\pi i)&=&C'\delta_{a+b,0}.
\label{eqn:no2}
\end{eqnarray}
Here $C,C'$ are constants depending on the normalization
of $Z_a(\beta),Z'_a(\beta)$. 
In addition, the symmetry relation \eqref{eqn:Lsym2} reduces to the following. 
\begin{equation}
Z'_a(\beta)Z_b(\beta-\frac{\pi i}{2})
=
Z_b(\beta^*)Z'_a(\beta^*-\frac{\pi i}{2}),
\qquad
\beta^*=\beta-\pi i (\xi+\frac{1}{2}). 
\label{eqn:ZZ4}
\end{equation}

One should note that the relations (\ref{eqn:ZZ1})-(\ref{eqn:ZZ3}) are much simpler
than those for $\Phi_\vep(\beta)$ and $\Psi^*_\vep(\beta)$. 
In the next section we will give a boson representation of $Z_a(\beta)$
and $Z'_a(\beta)$.

\vspace{5mm}
\noindent
{\it Remark.}
Let us denote the limit of the $L-$operators $L^{\pm}_{\varepsilon,\varepsilon'}(\zeta)$ by $L^{\pm}_{\varepsilon,\varepsilon'}(\beta)$. Then  the defining relation of the elliptic algebra (\ref{eqn:Rpm1})-(\ref{eqn:Lsym})
degenerate to  
\begin{eqnarray}
&&\tilde{R}^\pm_{12}(\beta_1-\beta_2)\Lpm{1}(\beta_1)\Lpm{2}(\beta_2)
=\Lpm{2}(\beta_2)\Lpm{1}(\beta_1)\tilde{R}^{*\pm}_{12}(\beta_1-\beta_2),
\label{eqn:tRpm1}\\
&&\tilde{R}^+_{12}(\beta_1-\beta_2-\frac{i\pi}{2})\Lp{1}(\beta_1)\Lm{2}(\beta_2)
=\Lm{2}(\beta_2)\Lp{1}(\beta_1)\tilde{R}^{*+}_{12}(\beta_1-\beta_2+\frac{i\pi}{2}),
\nonumber\\
&&\label{eqn:tRpm2}\\
&&\qdet L^+(\beta)\equiv
L^+_{++}(\beta+i\pi)L^+_{--}(\beta)-L^+_{-+}(\beta+i\pi)
L^+_{+-}(\beta)=1,
\label{eqn:tqdet}\\
&&L^-_{\vep\vep'}(\beta)=\vep\vep' L^+_{-\vep,-\vep'}(\beta-i\pi(\xi+\frac{1}{2})).
\label{eqn:tLsym}
\ena
These are similar to the defining relations of the quantum affine algebra
$U_q(\widehat{sl_2})$\cite{RS}.
However  the extra relation (\ref{eqn:tLsym}) indicates that our 
$L-$operators $L^\pm(\beta)$ have no Gauss decomposition. Our
resultant algebra is hence quite different from  $U_q(\widehat{sl_2})$ 
with $|q|=1$.
 
\subsection{Rational limit}

We next consider the rational limit $\xi\to\infty$ of the results in $\S 3.1$.
In this limit, we have 
\bea
&&\lim_{\xi\to \infty}\tilde{ R}^{*}(\beta)=S_R(\beta),\qquad 
\lim_{\xi\to \infty}\tilde{ R}(\beta)=R_R(\beta)
\ena
with
\bea
&&S_R(\beta)=S_{R,0}(\beta) 
\pmatrix{1&&&\cr
&\frac{\beta}{i\pi-\beta}&\frac{i\pi}{i\pi-\beta}&\cr
&\frac{i\pi}{i\pi-\beta}&\frac{\beta}{i\pi-\beta}&\cr
&&&1\cr},\\
&&S_{R,0}(\beta)=\frac{\Gamma(\frac{i\beta}{2\pi})
\Gamma(\frac{1}{2}-\frac{i\beta}{2\pi})}{\Gamma(-\frac{i\beta}{2\pi})\Gamma(\frac{1}{2}+\frac{i\beta}{2\pi})}
\ena
and 
\bea
&&R_R(\beta)=-S_R(-\beta).
\end{eqnarray}
Note that $S_R(\beta)$ and $R_R(\beta)$ are invariant
under the transformations (\ref{eqn:gauge1})-(\ref{eqn:gauge2}) 
and (\ref{eqn:gauge1*})-(\ref{eqn:gauge2*}) .
They coincide with the $S-$matrix of the $SU(2)$ invariant Thirring model and the $R-$matrix of the $XXX$ model, respectively.

Let  $L^{\pm}_{R\varepsilon,\varepsilon'}(\beta)$ be the rational limit of $L^{\pm}_{\varepsilon,\varepsilon'}(\beta)$. 
In this limit, the relations (\ref{eqn:tRpm1})-(\ref{eqn:tqdet}) are still hold with 
replacement $\tilde{R}^*(\beta) \to S_R(\beta)$, $\tilde{R}(\beta)\to R_R(\beta)$. However the last relation (\ref{eqn:tLsym}) is broken down
since the RHS loses its meaning. Hence the resultant algebra 
is the central extension of the Yangian double
${\cal D}Y(sl_2)$ at level one due to Reshetikhin and Semenov-Tian-Shanskii\cite{RS}.

\setcounter{section}{3}
\setcounter{equation}{0}
\def\vac{|{\rm vac}\rangle}

\section{Bosonization of the vertex operators}

We here consider the boson representation of the algebra 
(\ref{eqn:ZZ1})-(\ref{eqn:no2}).
 
\subsection{Trigonometric case}
Let us consider free bosons
$
a(t)\quad (t\in \R)
$
which satisfy\cite{JKM}
\begin{equation}
[a(t),a(t')]=\frac{\sinh{\frac{\pi t}{2}}\sinh\pi t
\sinh{\frac{\pi t(\xi+1)}{2}}}
{t\sinh\frac{\pi t\xi}{2}}\delta(t+t'). \label{eqn:3-1}\\
\end{equation}
We also use $a'(t)$ defined by
$$
a'(t)\sinh\frac{\pi t(\xi+1)}{2}=a(t)\sinh\frac{\pi t\xi}{2}.
$$
We consider the Fock space $\cal H$ generated by $\vac$ 
which satisfies
$$
a(t)\vac=0\quad\hbox{if}\quad t>0.
$$
We set
\begin{eqnarray*}
&V(\alpha)=:e^{i\phi(\alpha)}:,
\qquad 
&i\phi(\alpha)=\int^\infty_{-\infty}\frac{a(t)}{\sinh \pi t}e^{i\alpha t}dt,\\
&{\overline V}(\alpha)=:e^{-i\bar\phi(\alpha)}:,
\qquad
&i\bar\phi(\alpha)=\int^\infty_{-\infty}
\frac{a(t)}{\sinh\frac{\pi t}{2}}e^{i\alpha t}dt,\\
&V'(\alpha)=:e^{i\phi '(\alpha)}:,
\qquad
&i\phi '(\alpha)=-\int^\infty_{-\infty}\frac{a'(t)}{\sinh \pi t}
e^{i\alpha t}dt,\\
&{\overline V}'(\alpha)=:e^{-i\bar\phi '(\alpha)}:,
\qquad
&i\bar\phi '(\alpha)=-\int^\infty_{-\infty}
\frac{a'(t)}{\sinh\frac{\pi t}{2}}e^{i\alpha t}dt.
\end{eqnarray*}
In Appendix 1 in Ref.\cite{JKM}, one can find the list of the operator products of these operators.
%\footnote{ In the calculation, it should be understood that 
%when one 
%encounters an integral 
%\[
%\int_0^\infty F(t)dt
%\]
%which is divergent at $t=0$, we  regularize it  
%as the contour integral 
%\begin{equation}
%\int_C F(t)\frac{\log(-t)}{2\pi i}dt,  \label{eqn:3-00}
%\end{equation}
%with the contour $C$  
%$$
%\setlength{\unitlength}{1mm}
%\begin{picture}(70,20)
%\put(20,10){\oval(40,20)[l]}
%\put(20,10){$\cdot $}        %{$\bullet $}%
%\put(20,6){$0$}
%\put(20,20){\line(1,0){40}}
%\put(60,20){\vector(-1,0){20}}
%\put(20,0){\vector(1,0){20}}
%\put(40,0){\line(1,0){20}}
%\end{picture}
%$$
%.}

Let us define 
\begin{eqnarray}
Z_+(\beta)&=&V(\beta),\label{eqn:3-20}\\
Z_-(\beta)&=&\int_{C_1}\frac{d\alpha}{2\pi}
:e^{i\phi(\beta)-i\bar\phi(\alpha)}:f(\alpha-\beta),\label{eqn:3-2}\\
Z'_+(\beta)&=&V'(\beta),\label{eqn:3-30}\\
Z'_-(\beta)&=&\int_{C_2}\frac{d\alpha}{2\pi}
:e^{i\phi '(\beta)-i\bar\phi '(\alpha)}:f'(\alpha-\beta),\label{eqn:3-3}
\end{eqnarray}
where\footnote{
In order to make the expressions admit their rational limit,
we have changed the definition of  $f(\alpha)$ and $f'(\alpha)$ 
by adding extra constant factors
omitted in Ref.\cite{JKM}.
} 
\begin{eqnarray*}
f(\alpha)&=&c_1\sinh\frac{\pi}{\xi}
\ \Gamma\left(\frac{i\alpha}{\pi\xi}-\frac{1}{2\xi}\right)
\Gamma\left(-\frac{i\alpha}{\pi\xi}-\frac{1}{2\xi}\right),\qquad
c_1=\frac{e^{-(\gamma+\log\pi\xi)\frac{\xi}{\xi+1}}}{i\pi}
\label{func1},\\
f'(\alpha)&=&c_2\sinh\frac{\pi}{\xi+1}
\ \Gamma\left(\frac{i\alpha}{\pi(\xi+1)}+\frac{1}{2(\xi+1)}\right)
\Gamma\left(-\frac{i\alpha}{\pi(\xi+1)}+\frac{1}{2(\xi+1)}\right),
\label{func2}\\
&& c_2=\frac{e^{-(\gamma+\log\pi(\xi+1))
\frac{\xi+1}{\xi}}}{i\pi}.\nonumber
\end{eqnarray*}
Here the integration contours are chosen as follows. 
The contour $C_1$ is $(-\infty,\infty)$ except that the poles 
$
\beta-\frac{\pi i}{2}+n\pi\xi i\quad(n\in \Z_{\ge 0})
$ 
of 
$\Gamma(\frac{i(\alpha-\beta)}{\pi \xi}-\frac{1}{2\xi})$ are above
$C_1$ and the poles
$
\beta+\frac{\pi i}{2}-n\pi\xi i\quad(n\in \Z_{\ge 0})
$
of $\Gamma(-\frac{i(\alpha-\beta)}{\pi \xi}-\frac{1}{2\xi})$ are 
below $C_1$.
The contour $C_2$ is $(-\infty,\infty)$. The poles  
$
\beta+\frac{\pi i}{2}+n\pi(\xi+1)i\quad(n\in Z_{\ge 0})
$ 
of 
$\Gamma(\frac{i(\alpha-\beta)}{\pi(\xi+1)}+\frac{1}{2(\xi+1)})$ are above $C_2$
and the poles
$\beta-\frac{\pi i}{2}-n\pi(\xi+1)i\quad(n\in Z_{\ge 0})$
of $\Gamma(-\frac{i(\alpha-\beta)}{\pi(\xi+1)}+\frac{1}{2(\xi+1)})$ are below
$C_2$.
%The constants $c$ and $c'$ are given in Appendix 2. 

In \cite{JKM}, we proved the following statement.
\begin{prop}
The operators $Z_\pm(\beta)$ and $Z'_\pm(\beta)$
satisfy the commutation relations (\ref{eqn:ZZ1})-(\ref{eqn:ZZ3}),
the normalization conditions (\ref{eqn:no1}) and (\ref{eqn:no2}) as well as thesymmetry  relation (\ref{eqn:ZZ4}).
The constants $C,C'$ are 
given by
\begin{eqnarray*}
&&C=\left(\pi\xi c_1\sin\frac{\pi}{\xi}\
\Gamma(-\frac{1}{\xi})\right)^2g(-\pi i),
\\
&&C'=\left(\pi(\xi+1)c'_1\sin\frac{\pi}{\xi+1}\
\Gamma(\frac{1}{\xi+1})\right)^2
\lim_{\beta\rightarrow 0}\frac{g'(\beta+\pi i)}{\beta}.
\\
\end{eqnarray*}
\end{prop}

\subsection{Rational case}

Let us next consider the rational limit. 
We denote  the limit $\xi\to \infty$ 
of $a(t)$ and $a'(t)$ by $a_{R}(t)$ and $a'_{R}(t)$, respectively. They satisfy 
\bea
&&[a_R(t),a_R(t')]=\frac{\sinh{\frac{\pi t}{2}}\sinh\pi t
\ e^{\frac{\pi |t|}{2}}}
{t}\delta(t+t'). \label{eqn:r3-1}\\
&&a'_R(t)e^{\frac{\pi |t|}{2}}=a_R(t).
\ena
We denote  the corresponding limit of the boson fields, 
vertex operators and the Fock space by adding the suffix $R$. 
For example, $i\phi_R(\beta)=\lim_{\xi\to \infty}i\phi(\beta)=\int^\infty_{-\infty}\frac{a_R(t)}{\sinh \pi t}e^{i\alpha t}dt$.
Under these notations, the rational limit of the vertex operators $Z_{\pm}(\beta)$ and 
$Z'_{\pm}(\beta)$ are given by
\begin{eqnarray}
Z_{R+}(\beta)&=&V_R(\beta),\label{eqn:r3-20}\\
Z_{R-}(\beta)&=&\int_{C_{R1}}\frac{d\alpha}{2\pi}
:e^{i\phi_R(\beta)-i\bar\phi_R(\alpha)}:f_R(\alpha-\beta),\label{eqn:r3-2}\\
Z'_{R+}(\beta)&=&V'_R(\beta),\label{eqn:r3-30}\\
Z'_{R-}(\beta)&=&\int_{C_{R2}}\frac{d\alpha}{2\pi}
:e^{i\phi '_R(\beta)-i\bar\phi '_R(\alpha)}:f'_R(\alpha-\beta),
\label{eqn:r3-3}
\end{eqnarray}
where
\begin{eqnarray*}
f_R(\alpha)&=&f'_R(\alpha)=-{i\pi}e^{-\gamma}\frac{1}{\alpha^2+\frac{\pi^2}{4}}.
\end{eqnarray*}
The integration contours $C_{R1},\ C_{R2}$ should be  chosen as follows. 
The contour $C_{R1}$ is $(-\infty,\infty)$ except that the pole
$
\beta-\frac{\pi i}{2}
$ 
is above
$C_{R1}$ and the pole
$
\beta+\frac{\pi i}{2}
$
is below $C_{R1}$.
The contour $C_{R2}$ is $(-\infty,\infty)$. The pole  
$
\beta+\frac{\pi i}{2}
$ 
is  above $C_{R2}$
and the pole
$\beta-\frac{\pi i}{2}$ is  below $C_{R2}$.

In Appendix, we list all the operator products of the vertex operators 
$V_R(\alpha),\ \bar{V}_R(\alpha),\ V'_R(\alpha)$ and $\bar{V}'_R(\alpha)$.

After changing  the definition of $f(\alpha)$ and $f'(\alpha)$ to  (\ref{func1}) and (\ref{func2})
in the proof of Prop.3.1 and 3.3 in \cite{JKM}, whole arguments given there admit their rational limits. 
Hence we obtain 
\begin{prop}
The operators $Z_{R\pm}(\beta)$ and $Z'_{R\pm}(\beta)$
satisfy the rational limit of the 
commutation relations (\ref{eqn:ZZ1})-(\ref{eqn:ZZ3}) and 
the normalization conditions (\ref{eqn:no1})(\ref{eqn:no2}).
The corresponding constants $C_R,C_R'$ are 
given by
\begin{eqnarray*}
&&C_R=-\sqrt{2}e^{-\frac{3\gamma}{2}},
\\
&&C_R'=-i\sqrt{2}e^{-\frac{3\gamma}{2}}
\end{eqnarray*}
\end{prop}

Hence the operators $L^{\pm}_R(\beta)$
\begin{eqnarray}
L^+_{R\vep\vep'}(\beta)&=&\kappa_R \Psi^*_{R\vep'}(\beta)
\Phi_{R\vep}(\beta-\frac{\pi i}{2}),
\label{eqn:rMiki1}\\
L^-_{R\vep\vep'}(\beta)&=&\kappa_R \Phi_{\vep}(\beta)\Psi^*_{R\vep'}(\beta-\frac{\pi i}{2})
\label{eqn:rMiki2}
\end{eqnarray}
with (\ref{eqn:r3-20})-(\ref{eqn:r3-3}) and the gauge transformation (\ref{eqn:gauge1})-(\ref{eqn:gauge2*}) gives a boson representation of the central extension of the Yangian double at level one.
One should compare this with those in \cite{KI,Kho} . 

\setcounter{section}{4}
\setcounter{equation}{0}
\section{Form factor in the sine-Gordon theory}

%In this section, we derive a  
%solution of the difference equations \eqref{G1}-\eqref{G3} 
%algebraically, and obtain an integral formula for it.

Let us define the boost operator $H$ by 
\bea
&&H=\int_0^\infty dt \frac{t^2\sh\frac{\pi t(\xi+1)}{2}}{
\sh\frac{\pi t}{2}\sh{\pi t}\sh\frac{\pi t\xi}{2}}
a'(-t)a'(t),
\ena
which enjoys the property 
\bea
&&e^{\lambda H}a'(t)e^{-\lambda H}=e^{-\lambda t}a'(t).
\ena
Hence we have 
\bea
&&e^{\lambda H}X(\beta)e^{-\lambda H}
=X(\beta+i\lambda),\label{shift}
\ena 
for $X=V,\bar{V},V',\bar{V'}$.

Let us consider the operators 
$$
{\cal O}(\alpha_1,\cdots,\alpha_m)_{\vep_1,\cdots,\vep_{m}}=
Z'_{-\vep_1}(\alpha_1+\pi i)\cdots Z'_{-\vep_m}(\alpha_m+\pi i)
Z'_{\vep_m}(\alpha_m)\cdots Z'_{\vep_1}(\alpha_1)
$$
Using (\ref{eqn:ZZ3}), we have 
\begin{prop}
\bea
&&[{\cal O}(\alpha_1,\cdots,\alpha_m), Z_{\pm}(\beta)]
=0\qquad \forall\beta, \ \alpha_j\ (j=1,...,m).
\ena
\end{prop}

Now let us consider the following function.
\bea
&&F^{\cal O}(\beta_1,\cdots,\beta_{N})_{\mu_1,\cdots,\mu_{N}}
\nonumber\\
&&=\frac{\tr_{{\cal H}}(e^{-\lambda H}{\cal O}
(\alpha_1,\cdots,\alpha_m)
Z_{\mu_N}(\beta_N)Z_{\mu_{N-1}}(\beta_{N-1})\cdots
Z_{\mu_{1}}(\beta_{1}))}
{\tr_{{\cal H}}(e^{-\lambda H})}.
\label{eqn:tr}
\ena
By using the relations \eqref{eqn:ZZ1},
(\ref{shift}) and the cyclic property of trace, one can show that 
the function $F^{\cal O}(\beta_1,\cdots,\beta_{2n})_{\mu_1,\cdots,\mu_{2n}}$
satisfies the Smirnov's first axiom with the $S-$matrix of the sine-Gordon theory  and 
the following level zero deformed Knizhnik-Zamolodchikov equation.
\bea
&&F^{\cal O}(\beta_1,\cdots,\beta_{N}+i\lambda)_{\mu_1,\cdots,\mu_{N}}\nonumber\\
&&= S_{\mu_{1}\mu_N}^{\mu_{1}'\tau_1}(\beta_1-\beta_N)
S_{\mu_{2}\tau_1}^{\mu_{2}'\tau_2}(\beta_2-\beta_N)
\cdots S_{\mu_{N-1}\tau_{N-2}}^{\mu_{N-1}'\mu_{N}'}
(\beta_{N-1}-\beta_{N})\nonumber\\
&&\qquad\times 
F^{\cal O}(\beta_1,\cdots,\beta_{N})_{\mu_1',\cdots,\mu_{N}'}.
\ena
If one sets $\lambda=2\pi$, the $S-$matrix symmetry and the deformed KZ equation are equivalent to the Smirnov's first  and second  axioms.

In order to show that the function $F^{\cal O}(\beta_1,\cdots,\beta_{N})_{\mu_1,\cdots,\mu_{N}}$ satisfies the third axiom, let us consider  the relation
 (\ref{eqn:no1}). Applying  this to the product $Z_{\mu_N}(\beta_N)
Z_{\mu_{N-1}}(\beta_{N-1})$ in (\ref{eqn:tr}), one finds a simple pole at 
$\beta_N=\beta_{N-1}+\pi i$. In addition, noting the cyclic property of the trace 
and using (\ref{eqn:ZZ1}) and (\ref{shift}), one can change the order of $Z_{\mu_N}(\beta_N)$
and $ Z_{\mu_{N-1}}(\beta_{N-1})$ as follows.
\bea
&&S_{\mu_{N-1}\mu_{N-2}}^{\tau_{1}\tau_{N-2}'}(\beta_{N-1}-\beta_{N-2})
S_{\tau_{1}\mu_{N-3}}^{\tau_{2}\mu_{N-3}'}(\beta_{N-1}-\beta_{N-3})
\cdots S_{\tau_{N-2}\mu_{1}}^{\mu_{N-1}'\mu_{1}'}
(\beta_{N-1}-\beta_{1})\nonumber\\
&&\times
\frac{\tr_{{\cal H}}(e^{-\lambda H}{\cal O}
(\alpha_1,\cdots,\alpha_m)
Z_{\mu_{N-1}}(\beta_{N-1}+i\lambda)Z_{\mu_N}(\beta_N)
Z_{\mu_{N-2}}(\beta_{N-2})\cdots
Z_{\mu_{1}}(\beta_{1}))}
{\tr_{{\cal H}}(e^{-\lambda H})}.
\nonumber
\ena
The product $Z_{\mu_{N-1}}(\beta_{N-1}+i\lambda)Z_{\mu_N}(\beta_N)$
has a simple pole at $\beta_N=\beta_{N-1}+i\lambda-\pi i$. 
 Hence if one takes the limit $\lambda\to 2\pi$, the residue at the 
pole $\beta_N=\beta_{N-1}+\pi i$ is given by
\bea
&&2\pi i\ {\rm  res}\
F^{\cal O}(\beta_1,\cdots,\beta_{N})_{\mu_1,\cdots,\mu_{N}}
\nonumber \\
&&=C
F^{\cal O}(\beta_1,\cdots,\beta_{N-2})_{\mu'_1,\cdots,\mu'_{N-2}}
\delta_{\mu_N+\mu_{N-1},0}\nonumber\\
&&\times
\Bigl(
\delta_{\mu_1}^{\mu_1'}\cdots 
\delta_{\mu_{N-1}}^{\mu_{N-1}'}\nonumber\\
&&-
S_{\mu_{N-1}\mu_{N-2}}^{\tau_{1}\mu_{N-2}'}
(\beta_{N-1}-\beta_{N-2})
S_{\tau_{1}\mu_{N-3}}^{\tau_{2}\mu_{N-3}'}(\beta_{N-1}-\beta_{N-3})
\cdots S_{\tau_{N-2}\mu_{1}}^{\mu_{N-1}'\mu_{1}'}
(\beta_{N-1}-\beta_{1})
\Bigr) .\nonumber \label{residue}\\
\ena
The other cases $\beta_N=\beta_j+\pi i, \ j\leq N-2$ follows (\ref{residue}) and the $S-$matrix symmetry. Note that the charge conjugation matrix ${\cal C}_{\mu,\mu'}$ has been set $\delta_{\mu+\mu',0}$.

We hence conjecture that the function (\ref{eqn:tr}) provides  
a form factor of some local operator in the sine-Gordon theory.
In addition, it admit the rational limit $\xi\to \infty$. In this limit, the 
 function is expected to provide a form factor in the $SU(2)$
invariant Thirring model.

The boson realization discussed in $\S4$  allows the evaluation of the trace. We here  present only its final result. Setting 
$N=2n$, $\lambda=2\pi$,  $\mu_j=- $ for $ j=1,2,...,n$ and 
$\mu_j=+ $ for $j=n+1,...,2n$, 
 we get

\newpage
\bea
&&F^{\cal O}(\beta_1,\cdots,\beta_{2n})_{-,\cdots,-,+,\cdots,+}
\nonumber\\
&&=
\prod_{1\leq r<s\leq 2n}\zeta(\beta_r-\beta_s)
\prod_{j=1}^m\prod_{k=1}^{2n}
\frac{1}{
\sinh\left(
\frac{\pi i}{4}-
\frac{\beta_k-\alpha_j}{2}
\right)
}
\nonumber\\
&&\times\prod_{l=1}^n
\left(
\int_{ C_{\delta_l}}\frac{d\delta_l}{2\pi}
\right)
{\cal F}_{I-II}(\alpha_1,\cdots,\alpha_m;\delta_1,\cdots,\delta_n)\nonumber\\
&&\times \prod_{1\leq l<j\leq 2n}\varphi(\delta_l-\beta_j+\pi i )
\prod_{1\leq j<l\leq n}\varphi(\beta_j-\delta_l+\pi i )
\prod_{l=1}^{ n}\frac{\varphi(\beta_l-\delta_l+\pi i )}{\sinh\frac{1}{\xi}\left(
\beta_l-\delta_l-\frac{\pi i}{2} 
\right)
}\nonumber\\
&&\times\prod_{1\leq k<l\leq n}\sinh\frac{1}{\xi}(\delta_l-\delta_k-\pi i )
\sinh(\delta_l-\delta_k). \label{ffactor}
\ena
Here 
\bea
&&\zeta(\beta)=
\sinh\frac{\beta}{2}\exp\left(
\int_0^\infty \frac{dt}{t}\frac{\sin^2\left((
\beta+\pi i)\frac{t}{2}\right)\sinh(1-\xi)\frac{\pi t}{2}}
{
\sinh\frac{\pi t\xi}{2}\sinh\pi t\cosh\frac{\pi t}{2}
}
\right),\\
&&\varphi(\beta)
=\exp\left(
-2\int_0^\infty \frac{dt}{t}\frac{\sin^2\frac{\beta t}{2}
\sinh(1+\xi)\frac{\pi t}{2}}
{
\sinh\frac{\pi t\xi}{2}\sinh\pi t
}
\right),\\
&&{\cal F}_{I-II}(\alpha_1,\cdots,\alpha_m;\delta_1,\cdots,\delta_n)\nonumber\\
&&=\prod_{a\in A}\left(\int_{C_{\gamma_a}}\frac{d \gamma_a}{2\pi}\prod_{j=1}^m\frac{
1}{\cosh(\gamma_a-\alpha_j)}\right) 
\prod_{a'\in A'}\left(\int_{C_{\gamma'_{a'}}}\frac{d \gamma'_{a'}}{2\pi}\prod_{j=1}^m\frac{
1}{\cosh(\gamma'_{a'}-\alpha_j)}\right)\nonumber\\
&&\times
\prod_{a<b} \frac{\sinh(\gamma_a-\gamma_b)}
{\sinh\nu(\gamma_a-\gamma_b-\pi i)}
\prod_{a,a'} \frac{\sinh(\gamma'_{a'}-\gamma_a)}
{\sinh\nu(\gamma'_{a'}-\gamma_a-\pi i)}
\prod_{a'<b'} \frac{\sinh(\gamma'_{a'}-\gamma'_{b'})}
{\sinh\nu(\gamma'_{a'}-\gamma'_{b'}-\pi i)}\nonumber\\
&&\times
\prod_{a\in A}\left( \prod_{a<j}{\sinh\nu(\gamma_a-\alpha_j+\frac{\pi i}{2})} \prod_{j<a}{\sinh\nu(\alpha_j-\gamma_a+\frac{\pi i}{2})}\right)
\nonumber \\
&&\times\prod_{a'\in A'}\left( \prod_{j<a'}{\sinh\nu(\gamma'_{a'}-\alpha_j-\frac{\pi i}{2})} \prod_{a'<j}{\sinh\nu(\alpha_j-\gamma'_{a'}+\frac{3\pi i}{2})}\right)
\nonumber\\
&&\times\frac{\prod_{k=1}^{2n}\left(
\prod_{a\in A}\sinh\frac{1}{2}(\beta_k-\gamma_{a})
\prod_{a'\in A'}\sinh\frac{1}{2}(\beta_k-\gamma'_{a'})\right)
\prod_{j=1}^m\prod_{l=1}^n\sinh(\delta_l-\alpha_j)}
{
\prod_{l=1}^{n}\left(
\prod_{a\in A}\cosh(\delta_l-\gamma_{a})
\prod_{a'\in A'}\cosh(\delta_l-\gamma'_{a'})\right)
}.
\nonumber\\
\label{fiii}
\ena
In (\ref{fiii}), $\nu=1/(\xi+1)$,  $A=\{j|1\leq j\leq m, \varepsilon_j=-\}$ and $A'=\{j|1\leq j\leq m, \varepsilon_j=+\}$. 
The integration contours are determined from $C_1$ and $C_2$ in (\ref{eqn:3-2}) and (\ref{eqn:3-3}) and the convergence region of the operator products of the vertex operators $V, \bar V, V', \bar V'$ listed in Appendix of Ref.\cite{JKM}. We chose them as follows.

\noindent
The contour $C_{\delta_l}$ is $(-\infty, \infty)$ except that
the poles at 
\begin{eqnarray*}
&&\beta_j-\frac{\pi i }{2}+2\pi i n_1+i\pi \xi n_2\ 
 (l<j),\ 
\beta_j+\frac{3\pi i }{2}+2\pi i n_1+i\pi  \xi n_2 \ (j\leq l), \\
&&\beta_l-\frac{\pi i }{2}+\pi i \xi n_2, \ 
\gamma_{a}-\frac{\pi i}{2}+i\pi n_1, \  \gamma'_{a'}-\frac{\pi i}{2}+i\pi n_1
\end{eqnarray*}
$(n_1, n_2\in \Z_{\geq 0})$ are above $C_{\delta_l}$ and the poles  at 
\begin{eqnarray*}
&&\beta_j-\frac{3\pi i }{2}-2\pi i n_1-i\pi  \xi n_2\ (l<j),\  
\beta_j+\frac{\pi i }{2}-2\pi i n_1-i\pi \xi n_2\  (j\leq l), \\ 
&&\gamma_{a}-\frac{3\pi i}{2}-i\pi n_1, \  \gamma'_{a'}-\frac{3\pi i}{2}-i\pi n_1
\end{eqnarray*}
$(n_1, n_2\in \Z_{\geq 0}) $ are below $C_{\delta_l}$.

\noindent
The contour $C_{\gamma_a}$ is $(-\infty, \infty)$ except that  
the poles at 
\begin{eqnarray*}
&&\alpha_j+\frac{\pi i }{2}+i\pi  n\  (a\leq j),\
\alpha_j+\frac{3\pi i }{2}+i\pi  n\  (j< a),\\
&& \gamma_b+\pi i+(n+1) \frac{\pi i }{\nu} \ (a<b),\ 
\gamma_b-\pi i+(n+1) \frac{\pi i }{\nu} \ (b<a),\\
&&\gamma_{a'}-\pi i+(n+1) \frac{\pi i }{\nu},\
\delta_{l}+\frac{3\pi i }{2}+i\pi n 
\end{eqnarray*}
 $ (  n\in \Z_{\geq 0})$ are above $C_{\gamma_a}$ and the poles  at 
\begin{eqnarray*}
&&\alpha_j-\frac{3\pi i }{2}-i\pi  n\ (a<j),\
\alpha_j-\frac{\pi i }{2}-i\pi  n \ (j\leq a),\\
&&\gamma_b+\pi i-(n+1) \frac{\pi i }{\nu} \ (a<b),\
\gamma_b-\pi i-(n+1) \frac{\pi i }{\nu} \ (b<a),\\ 
&&\gamma_{a'}-\pi i-(n+1) \frac{\pi i }{\nu},\
\delta_{l}+\frac{\pi i }{2}-i\pi n \
\end{eqnarray*}
$ (  n\in \Z_{\geq 0})$ are below $C_{\gamma_a}$.

\noindent
The contour $C_{\gamma'_{a'}}$ is $(-\infty, \infty)$ except that
the poles at 
\begin{eqnarray*}
&&\alpha_j+\frac{5\pi i }{2}+i\pi  n\ (a'<j),\
\alpha_j+\frac{3\pi i }{2}+i\pi  n\ (j\leq a'),\\
&&\gamma_{a}+\pi i+(n+1) \frac{\pi i }{\nu} ,\
 \gamma_{b'}+\pi i+(n+1) \frac{\pi i }{\nu} \ (a'<b'),\\
&&\gamma_{b'}-\pi i+(n+1) \frac{\pi i }{\nu} \ (b'<a'),\
\delta_{l}+\frac{3\pi i }{2}+n\pi i \ 
\end{eqnarray*}
$(  n\in \Z_{\geq 0})$ are above $C_{\gamma'_{a'}}$ and the poles  at 
\begin{eqnarray*}
&&\alpha_j+\frac{\pi i }{2}-i\pi  n \ (a'\leq j),\
\alpha_j-\frac{\pi i }{2}-i\pi  n\  (j< a'),\\
&&\gamma_{a}+\pi i-(n+1) \frac{\pi i }{\nu},\
\gamma_{b'}+\pi i-(n+1) \frac{\pi i }{\nu} \ (a'<b'),\\
&&\gamma_{b'}-\pi i-(n+1) \frac{\pi i }{\nu} \ (b'<a'),\ 
\delta_{l}+\frac{\pi i }{2}-n\pi i \  
\end{eqnarray*}
$(  n\in \Z_{\geq 0})$ are below $C_{\gamma_{a'}}$.

In (\ref{ffactor}), we have omitted a constant factor, which depends on $\xi$. 
When one considers the rational limit, this factor should be properly considered. The  rational limit yields the formula
in $\S10.4$ in Ref.\cite{JM} obtained as the rational limit of  the form factor in the
$XXZ$ model in the antiferromagnetic regime and  (7.14) in Ref.\cite{KLP}
with $\hbar=-i\pi$ as a special case $m=1$.  

The detail of the  calculation and the relation with the Smirnov's
integral formula will be discussed  in elsewhere.

\vspace{5mm}
\noindent
{ \bf Acknowledgments}\newline
The author would like to thank
the organizers of the workshop
Yvan Saint-Aubin, Luc Vinet and Mo-Lin Ge 
for kind invitation and hospitality. 
He is also grateful to Michio Jimbo and  Tetsuji Miwa for 
collaboration in the work \cite{JKM} and for stimulating discussions.

%\appendix
\setcounter{equation}{0}
\newcommand{\im }{{\rm Im}}

\section{Appendix}

Here we list the formulas of the form
$$
X(\beta _1)Y(\beta _2) = C_{X,Y}(\beta _2-\beta _1)
:X(\beta _1)Y(\beta_2):
$$
where $X,Y=V_R,\bar{ V}_R, V_R', \bar{ V'}_R$
and $C_{X,Y}(\beta )$ is a meromorphic function on
$\C .$
The equality $C_{X,Y}=C_{Y,X}$ is valid in all cases. 
The whole expressions are obtained by taking the limit $\xi\to \infty$
in those listed in Ref.\cite{JKM}.

\begin{eqnarray}
& &V_R(\beta _1)V_R(\beta _2)=g_R(\beta _2-\beta _1)
   :e^{i\phi_R(\beta _1)+i\phi_R(\beta _2)}:\quad (\im (\beta _2-\beta _1)<0)
\label{eqn:A-1}
\\
& &g_R(\beta )=\sqrt{2\pi }e^{\gamma/2}
  \frac{\Gamma(\frac{1}{2}+\frac{i\beta}{2\pi})}
   {\Gamma (\frac{i\beta }{2\pi})}
\nonumber \\
& &V_R(\beta _1)\bar V_R(\beta _2)=w_R(\beta _2-\beta _1)
   :e^{i\phi_R(\beta _1)-i\bar \phi_R(\beta _2)}:\quad (\im (\beta _2-\beta _1)
    <-\frac{\pi}{2})
\label{eqn:A-2}\\
& &w_R(\beta )=
  \frac{e^{-\gamma}}
       {i (\beta+\frac{\pi i}{2})}
   =\frac{1}{g_R(\beta+\frac{\pi i}{2})g_R(\beta-\frac{\pi i}{2})}
\nonumber\\
& &\bar V_R(\beta _1)\bar V_R(\beta _2)=\bar g_R(\beta _2-\beta _1)
   :e^{-i\bar \phi_R(\beta _1)-i\bar \phi_R(\beta _2)}:\quad 
   (\im (\beta _2-\beta _1)<-\pi)
\label{eqn:A-4}
\\
& &\bar g_R(\beta )=
   -e^{2\gamma}\beta(\beta+\pi i)
  =\frac{1}{w_R(\beta+\frac{\pi i}{2})w_R(\beta-\frac{\pi i}{2})}
\nonumber \\
& &V_R'(\beta _1)V_R'(\beta _2)=g_R'(\beta _2-\beta _1)
   :e^{i \phi_R '(\beta _1)+i \phi_R '(\beta _2)}:\quad 
   (\im (\beta _2-\beta _1)<\pi)
\\
& &g_R'(\beta )=
\sqrt{2\pi }e^{\gamma/2}
  \frac{\Gamma(1+\frac{i\beta}{2\pi})}
   {\Gamma (\frac{1}{2}+\frac{i\beta }{2\pi})}
\nonumber \\
& &V_R'(\beta _1)\bar V_R'(\beta _2)=w_R'(\beta _2-\beta _1)
   :e^{i \phi_R '(\beta _1)-i \bar \phi_R '(\beta _2)}:\quad 
   (\im (\beta _2-\beta _1)<\frac{\pi}{2})
\\
& &w_R '(\beta )=
  \frac{e^{-\gamma}}
       {i (\beta-\frac{\pi i}{2})}
=\frac{1}{g_R'(\beta+\frac{\pi i}{2})g_R'(\beta-\frac{\pi i}{2})}
\nonumber \\
& &\bar V_R'(\beta _1)\bar V_R'(\beta _2)=\bar g_R'(\beta _2-\beta _1)
   :e^{-i \bar \phi_R '(\beta _1)-i \bar \phi_R '(\beta _2)}:\quad 
   (\im (\beta _2-\beta _1)<0)
\\
& &\bar g_R'(\beta )=
   -e^{2\gamma}\beta(\beta-\pi i)
 =\frac{1}{w_R'(\beta+\frac{\pi i}{2})w_R'(\beta-\frac{\pi i}{2})}
\nonumber \\
& & V_R(\beta _1) V_R'(\beta _2)=h(\beta _2-\beta _1)
   :e^{i \phi_R (\beta _1)+i \phi_R '(\beta _2)}:\quad 
   (\im (\beta _2-\beta _1)<\frac{\pi}{2})
\label{A.5} \\
& &h(\beta )=
         \frac{\Gamma (\frac{i\beta}{2\pi}+\frac{1}{4})}
       {\Gamma (\frac{i\beta}{2\pi}+\frac{3}{4})}
  e^{-\frac{1}{2}(\gamma +\log (2\pi ))}
\nonumber \\
& & V_R(\beta _1)\bar V_R'(\beta _2)=i(\beta _2-\beta _1)e^{\gamma }
   :e^{i \phi_R (\beta _1)-i \bar \phi_R '(\beta _2)}:\quad 
   (\im (\beta _2-\beta _1)<0)
\label{A.6} \\
& & \bar V_R(\beta _1) V_R'(\beta _2)=i(\beta _2-\beta _1)e^{\gamma }
   :e^{-i \bar \phi_R (\beta _1)+i \phi_R '(\beta _2)}:\quad 
   (\im (\beta _2-\beta _1)<0)
\label{A.7} \\
& & \bar V_R(\beta _1)\bar V_R'(\beta _2)=
 -\frac{e^{-2\gamma }}{(\beta _2-\beta _1)^2+\frac{\pi ^2}{4}}
   :e^{-i \bar \phi_R (\beta _1)-i \bar \phi_R '(\beta _2)}:\quad 
   (\im (\beta _2-\beta _1)<-\frac{\pi}{2})
\nonumber\\\label{A.8}
\end{eqnarray}

We set 
\begin{equation}
S_{R0}(\beta ) = \frac{g_R(-\beta )}{g_R(\beta )},
\qquad
R_{R0}(\beta ) = \frac{g_R'(-\beta )}{g_R'(\beta )}. 
\label{eqn:S0}
\end{equation}
The following relations are valid. 
\begin{eqnarray}
& &\frac{w_R(\beta)}{w_R(-\beta)}=
  -\frac{\beta -\frac{\pi i}{2}}
        {\beta +\frac{\pi i}{2}}, 
\label{eqn:A-3}
\\
& &\frac{\bar g_R(\beta)}{\bar g_R(-\beta)}=
  \frac{\beta +\pi i}
       {\beta -\pi i},
\\
& &\frac{w_R'(\beta)}{w_R'(-\beta)}=
  -\frac{\beta +\frac{\pi i}{2}}
       {\beta -\frac{\pi i}{2}},
\\
& &\frac{\bar g_R'(\beta)}{\bar g_R'(-\beta)}=
  \frac{\beta - \pi i}
       {\beta +\pi i},
\\
& &\frac{h(\beta)}{h(-\beta)}=
  -\frac{\sinh (\frac{\beta}{2} - \frac{\pi i}{4})}
       {\sinh (\frac{\beta}{2} + \frac{\pi i}{4})}.
\end{eqnarray}

%\bibliographystyle{unsrt}
%\bibliography{q}

\end{document}